\documentclass[a4paper,fleqn,usenatbib]{mnras}

\usepackage[T1]{fontenc}
\usepackage{ae,aecompl}

\usepackage{graphicx}	% Including figure files
\usepackage{amsmath}	% Advanced maths commands
\usepackage{amssymb}	% Extra maths symbols
\usepackage{pdflscape}

\newcommand{\msun}{M$_{\odot}$} % kilometres per second
\newcommand{\kms}{\,km\,s$^{-1}$} % kilometres per second
\newcommand{\ha}{H$\alpha$} %Halpha line
\title[Accretion and Outflow in V404 Cyg]{Accretion and Outflow in V404 Cyg}

\author[J. Casares et al.]{J. Casares$^{1,2}$\thanks{E-mail: jorge.casares@iac.es}, 
T. Mu\~noz-Darias$^{1,2}$, D. Mata S\'anchez$^{3}$, P.A. Charles$^{4,5}$,  
\newauthor M.A.P. Torres$^{1,2,6}$, M. Armas Padilla$^{1,2}$, R.P. Fender$^{4}$ and J. Garc\'ia-Rojas$^{1,2}$ 
\\
$^{1}$Instituto de Astrof\'\i{}sica de Canarias, 38205 La Laguna, Tenerife, Spain\\
$^{2}$Departamento de Astrof\'isica, Universidad de La Laguna, E-38206 La Laguna, Tenerife, Spain\\
$^{3}$Jodrell Bank Centre for Astrophysics, School of Physics and Astronomy, The University of Manchester, Manchester M13 9PL, UK\\
$^{4}$Department of Physics, Astrophysics, University of Oxford, Denys Wilkinson Building, Keble Road, Oxford OX1 3RH, UK\\
$^{5}$Department of Physics and Astronomy, University of Southampton, Southampton SO17 1BJ, UK\\
$^{6}$SRON, Netherlands Institute for Space Research, Sorbonnelaan 2, 3584 CA, Utrecht, The Netherlands \\
}

\date{Accepted XXX. Received YYY; in original form ZZZ}

\pubyear{2019}

\begin{document}
\label{firstpage}
\pagerange{\pageref{firstpage}--\pageref{lastpage}}
\maketitle

\begin{abstract}
We study the optical evolution of the 2015 outburst in V404 Cyg, with emphasis on the peculiar 
nebular phase and subsequent decay to quiescence. From the decay timescale of the Balmer 
emission associated with the nebula we measure an outflow mass 
$M_{\rm wind}\simeq4\times10^{-6}$ \msun. Remarkably, this is $\sim$100 times larger than the 
accreted mass and $\sim$10\% of the total mass stored in the disc. The wind efficiency must 
therefore be significantly larger than previous estimates for black hole transients, suggesting that 
radiation pressure (in addition to other mechanisms such as Compton-heating) plays a key role 
in V404 Cyg. In addition, we compare the evolution of the 2015 and 1989 outbursts and find clear 
similarities (namely a large luminosity drop $\sim$10 d after the X-ray trigger, followed by a brief 
nebular phase) but also remarkable differences in decay timescales and long-term evolution of 
the \ha~profile. In particular, we see evidence for a rapid disc contraction in 2015, consistent with 
a burst of mass transfer. This could be driven by the response of the companion to hard X-ray 
illumination, most notably during the last gigantic (super-Eddington) flare on 25 June 2015. 
We argue that irradiation and consequential disc wind are key factors to understand the different 
outburst histories in 1989 and 2015. In the latter case, radiation pressure may be responsible for 
the abrupt end of the outburst through depleting inner parts of the disc, thus quenching accretion 
and X-ray irradiation. We also present a refined orbital period and updated ephemeris. 
\end{abstract}


%%%%%%%%%%%%%%%%%%%%%%%%%%%%%%%%%%%%%%%%%%%%%%%%%%

%%%%%%%%%%%%%%%%% BODY OF PAPER %%%%%%%%%%%%%%%%%%

\section{Introduction}
\label{intro}

Black holes (BH) are key to understand extreme processes in the Universe, such as accretion 
and outflows. In particular, stellar-mass BHs within X-ray binaries present us with the opportunity 
to study these processes on human timescales \citep{mcclintock06,charles06,fender06}
and, therefore, often in greater detail than supermassive BHs (although see e.g. \citealt{eht19}). 
The large majority of Galactic stellar-mass BHs are found in X-ray transients (XRTs), a subclass of 
X-ray binaries that exhibit violent outbursts, with 18 (dynamically) confirmed cases and more than 60 candidates  
\citep{casares-jonker14,corral16}. 

Among BH XRTs, V404 Cyg, the optical counterpart of the {\it Ginga} source GS 2023+338, 
stands out as a benchmark source for a number of reasons. 
It was the first BH  with a dynamical mass comfortably exceeding that of  
standard neutron stars \citep{casares92}. Its companion star showed   
an anomalously large $^7$Li abundance \citep{martin92}, later proved to be a common feature 
of XRTs \citep{martin96}.  
It exhibited unprecedented X-ray variability during the 1989 outburst, with evidence for 
heavy line-of-sight absorption \citep{oosterbroek97,zycki99}.  
It is also the first BH XRT with a precise parallax distance \citep{miller-jones09} and 
the most luminous in quiescence at X-ray and radio energies 
\citep{kong02,gallo05}. Finally, it was the first BH XRT to show large flickering 
variability during quiescence in optical \citep{wagner92,zurita03}, near-IR \citep{sanwal96,zurita04}, 
X-rays \citep{wagner94,hynes04,bernardini14,rana16} and radio frequencies \citep{miller-jones08,rana16,plotkin19}.  

The renewed 2015 activity thus offered the opportunity to study another V404 Cyg outburst with 
modern instrumentation and a number of remarkable results were promptly produced. 
A non-exhaustive list includes: 
\begin{itemize}
\item[](1) the presence of long time-scale (correlated) optical/X-ray oscillations, suggestive of 
mass accretion disruption \citep{kimura15,alfonso-garzon18} 
\item[](2) evidence for a sustained disc outflow in the optical and X-rays  \citep{munoz16, king15}  
\item[](3) large X-ray obscuration (approaching Compton-thick at times) by the inner disc, 
possibly puffed-up by super-Eddington luminosity \citep{sanchez17, motta17a, motta17b, walton17}. 
\item[](4) detection of e$^-$ e$^+$ pair annihilation and $\gamma$-ray excess, in support for plasmoid 
ejections in a jet \citep{siegert16, loh16,piano17} 
\item[](5) evidence for discrete relativistic ejections in a precessing jet and   
strong disc-jet coupling \citep{miller-jones19, plotkin17,  tetarenko17}						
\item[](6) a tight constraint on the size of the jet launching/collimation region \citep{gandhi17}
\item[](7) a measurement of the magnetic field in the accretion disc corona \citep{dallilar18}.  
\end{itemize}

After 12 days of frantic activity, and following a gigantic 8-hr flare that peaked above 50 Crab in 
hard X-rays, the outburst suddenly ended \citep{sanchez17, jourdain17}. The rapid luminosity 
drop that followed revealed a brief nebular phase never seen before in any other BH XRT 
(\citealt{munoz16}; see also \citealt{rahoui17}). 
In \cite{mata18} we have presented a detailed spectroscopic study of 
the main 2015 outburst and its sequel a few months later (hereafter referred to as {\it mini-outburst}; \citealt{munoz17}), including emission line diagnostics specifically designed to detect outflow signatures.  
Here in this paper we focus on the properties of the unique nebular episode and the subsequent 
decay to quiescence. In particular, we exploit the evolution 
of the strongest emission line, \ha, to gain insights into the physics of the outflow and trace the response 
of the accretion disc to the outburst. 
In addition, we compare the evolution of the 2015 and 1989 outbursts and revisit the fundamental binary 
parameters, presenting a new updated ephemeris.   

\section{Observations and data reduction}
\label{sec:obs}

We use optical spectra of V404 Cyg collected between 15 June 2015 and 4 Jul 2016, thus 
covering the main outburst, the mini-outburst and early quiescence. 
The database consists of 
639 \ha~spectra already reported in \cite{munoz16,munoz17} and \cite{mata18}. 
These were collected with various instruments and resolutions, spanning between 60-300 \kms. 
Most spectra have been flux calibrated through observations of a spectrophotometric standard star, 
with slit-loss corrections provided by a comparison star in the slit (see \citealt{mata18}). 
Full details on the different instrument layouts can be found in Tables A1-A3 of \cite{mata18}. 

In addition, we include  
eight more quiescent spectra gathered between 21 Sept 2016 and 7 Nov 2018. 
These spectra were obtained with the Andalucia Faint Object Spectrograph and
Camera (ALFOSC) attached to the 2.5m Nordic Optical Telescope (NOT). We employed 
grism \#17 which covers the \ha~region at  high-resolution (R$\sim6000$). A log of the 
new quiescent observations is presented in Table~\ref{tab:tab1}.  

 \begin{table}
	\centering
	\caption{New NOT quiescent spectra}
	\label{tab:tab1}
	\begin{tabular}{ccccc}
		\hline
   Date & Instrument  & Coverage & Resolution & Texp \\ 
 & & (\AA) & \kms & \\
 		\hline
21/09/2016 & ALFOSC+GR17 &  6300-6900 & 40 & 2x2600s  \\
20/09/2017 & ,, &  ,, & 55 & 2x1800s  \\
08/08/2018 & ,, &  ,, & 55 & 2x1800s  \\
07/11/2018 & ,, &  ,, & 55 & 2x1800s  \\
\hline
	\end{tabular}
\end{table}

The recent spectroscopy was processed with standard debiasing, 
flat-field correction and optimal extraction using STARLINK/PAMELA routines \citep{marsh89}. 
Observations of Ne arc lamps were employed to derive pixel-to-wavelength calibrations through 4th order 
polynomial fits to more than 9 lines across the entire wavelength range. These yield a mean 
spectral dispersion of 0.29 \AA~pix$^{-1}$ and $<$0.047 \AA~rms.   
The accuracy of the wavelength calibration was checked against the \ha~sky line and was always 
found to be within 1 \kms. 

\section{Long-term view of the 2015 outburst}
\label{sec:2015}

\begin{figure}
	\includegraphics[angle=0,width=\columnwidth]{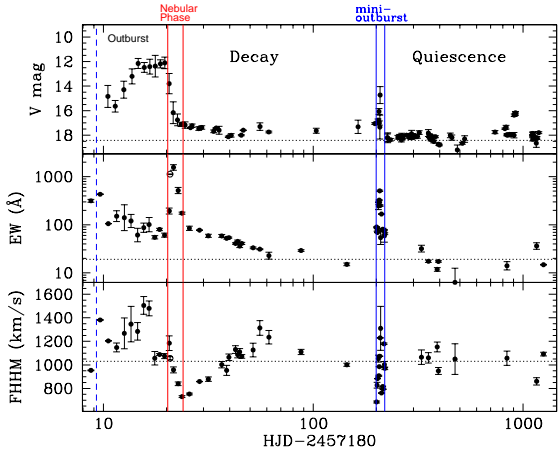}
    \caption{Long-term evolution of the 2015 outburst of V404 Cyg and its return to quiescence.  
    The top panel is the AAVSO V-band light curve, while the middle and bottom panels show the 
    \ha~EW and FWHM, respectively. Open circles represent data points obtained from  
   Table 2 in Rahoui et al. (2017). The blue dashed vertical line indicates the time of the X-ray trigger. 
   Dotted horizontal lines provide reference quiescence values from 
   Casares et al (1993) and Casares (2015), see text.
    }
    \label{fig:fig1}
\end{figure}

Figure~\ref{fig:fig1} presents the evolution of the \ha~EW and FWHM between 
June 2015 and Dec 2018, compared to the AAVSO\footnote{American Association of Variable Star  Observers; https://www.aavso.org} V-band light curve. 
Data points represent 1 day averages, with uncertainties indicating one standard deviation in the distribution 
of individual values. Therefore, except for a few cases with poor statistics and/or signal-to-noise, the error 
bars reflect intrinsic variability within 1 day bins. Following \cite{mata18} we tentatively define the limits 
of the nebular phase by the times when EW(\ha)$>$100 \AA~and these are represented by red vertical 
lines on the figure. The time of the {\it Swift} X-ray trigger \citep{barthelmy15} and the limits of the 
subsequent mini-outburst 
are also indicated by blue vertical lines. 
To guide the eye, horizontal dotted lines mark the V-mag and EW/FWHM values measured 
in quiescence, as reported in \cite{casares93} and \cite{casares15} respectively. 

Figure~\ref{fig:fig1} clearly shows how the nebular phase is triggered by a 
dramatic luminosity fall (5 optical mags in $\sim$2 days) that takes place on 26 June 2015 
(MJD=57200).  
We observe that, at variance with the more erratic behaviour seen in outburst, both 
the EW and FWHM follow 
smooth trends throughout the nebular phase:  
the EW rises rapidly (peaking at $\sim$2000 \AA) followed by a somewhat slower  
decay, while the FWHM shows a steady decrease from $\sim$1200 \kms~to $\sim$700 \kms. 
The EW evolution is explained because the irradiated continuum drops faster than the 
\ha~flux \citep{mata18}. 
On the other hand, the decrease in FWHM is caused by the 
dimming of the nebular contribution to the \ha~profile (see Sect.~\ref{sec:nebular}). 

After the 
nebular phase, the V-band flux 
falls off gradually\footnote{An exponential fit yields an e-folding timescale of 30 d.} 
until reaching a standstill level $\sim$0.5-1 mag above quiescence.  
Meanwhile, the EW continues its decline towards quiescence but at  
a flatter rate than during the nebular decay. On the other hand, the FWHM reverses its  
downward drift and starts tracing a lengthy oscillation around the quiescent value. 
After the mini-outburst episode, the V-mag and the two \ha~line parameters settle down at 
quiescent levels, although with significant residual variability that is characteristic of quiescent XRTs 
(see e.g. \citealt{hynes02}). 
Note that this is in contrast with X-ray results which, based on spectral softening, 
indicate the inner disc entered quiescence sometime between 23 July and 5 Aug 
(MJD 57227-57240), with $L_{\rm X}<10^{-5.6} L_{\rm Edd}$ \citep{plotkin17}. 
The optical evolution, therefore, suggests the outer disc 
is still responding to the outburst long after accretion onto the BH has stopped. 
Henceforth we decide to refer to the time between the nebular phase and the mini-outburst episode 
as {\it decay} while we call the period following the mini-outburst as {\it quiescence}.  

 \section{The Nebular phase and decay}
\label{sec:nebular} 

\begin{figure}
	\includegraphics[angle=-90,width=\columnwidth]{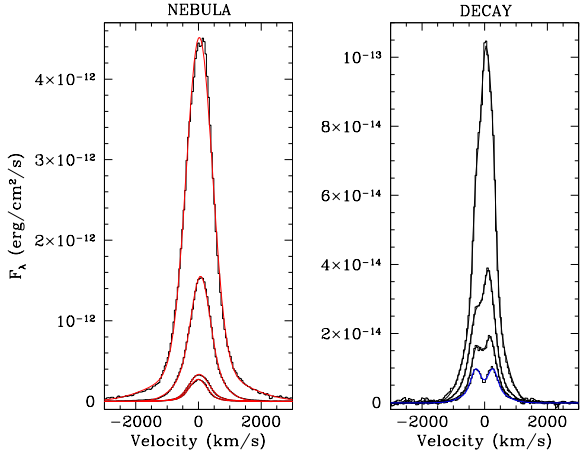}
    \caption{\ha~line profiles through the nebular phase and decay. The left panel displays 
    (from top to bottom) spectra obtained on days MJD 57200-3, when the nebula emission is entirely 
    dominant. 
    The right panel shows the gradual appearance of the accretion disc component under the nebula, 
    with spectra from MJD 57205, 57208, 57211 and 57225 (from top to bottom). 
    Red lines depict the best fits obtained with a nebula model, the blue line a disc model fit and 
    black lines fits from a combined disc+nebula model 
    (see text for details). } 
    \label{fig:fig2}
\end{figure}

To study the evolution of V404 Cyg through the nebular phase and beyond we focus on the 
 \ha~recombination line. We have produced nightly averages of flux calibrated spectra obtained 
 from MJD 57200 onwards. The spectra have been corrected for reddening using 
 $A_{\rm V}=4.0$ \citep{casares93,hynes04,miller-jones09}, 
 together with a mean Galactic extinction law and $R_{V}=A_{V}/E(B-V)=3.1$ \citep{howarth83}.
  A close inspection of the \ha~profiles 
reveals the fading of the strong broad single-peaked  component from the nebula,  
while an underlying double peak profile, associated with the accretion disc, gradually emerges   
(see Fig~\ref{fig:fig2}).  
In order to isolate these two contributions we decided to perform multi-Gaussian fits to the line profiles. 
We restrict ourselves to data obtained with velocity resolution $\leq$150 \kms~so as to better 
deconvolve the two 
components. Prior to the fit, the multi-Gaussian models were degraded to the resolution of the 
data through convolution with the appropriate instrumental profile. 

\begin{figure}
	\includegraphics[angle=-90,width=\columnwidth]{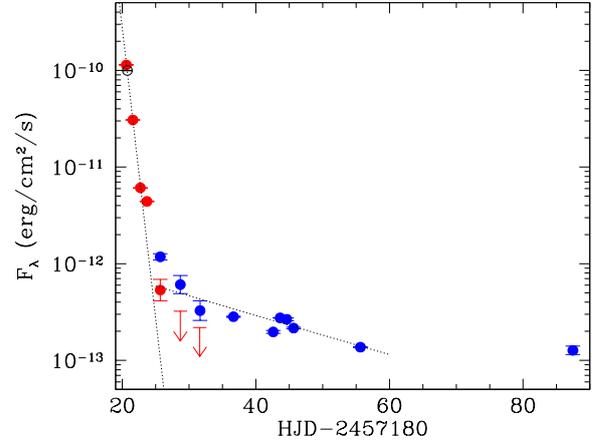}
    \caption{Decay curves of the nebular (red) and disc (blue) \ha~line components. The open black circle 
     corresponds to the \ha~flux quoted in Table 2 of Rahoui et al. (2017), dereddened by $A_{V}=4.0$. 
     Black dotted lines  represent the best exponential fits to the points. 
     The two last blue points have consistent fluxes, an indication that the disc \ha~flux has stopped  
     its decay on MJD $\sim$57240. Consequently, we have excluded the latest blue point 
     from the exponential fit. } 
    \label{fig:fig3}
\end{figure}

The broad nebular component is overwhelmingly dominant on days MJD 57200-57203 
and can be described by a two-Gaussian model, consisting of a broad base and a narrow core 
(left panel in Fig~\ref{fig:fig2}). 
On the other hand, \ha~lines from MJD 57216 onwards are characterized by a pure accretion disc profile. 
These have been modelled using three Gaussians: a symmetric double peak plus a broad base, fixed at the 
position of the double peak centroid. Finally, profiles obtained between MJD 57205 and 57211 show the 
incremental appearance of the accretion disc 
under the nebula contribution. In this case we have added an extra Gaussian to the accretion 
disc model, so as to account for the residual nebula emission. Figure~\ref{fig:fig2} shows a 
sample of \ha~fits while Fig~\ref{fig:fig3} plots the time evolution of the \ha~flux for the nebular 
and disc components, as derived from the multi-Gaussian fits\footnote{Note that, although the 
nebula component is loosly constrained on MJD 57205-57211 by the four Gaussian model, this 
has no impact on the outflow mass (see Sect.~\ref{sec:mass}) since the decay timescale of the 
nebula flux is already determined from observations on the early days of the nebular phase.}. 
The resulting fluxes are listed in Table~\ref{tab:tab2}. For comparison, the table 
also provides fluxes for H$\beta$ and the HeII $\lambda$4686+Bowen blend, obtained by 
direct integration of the line profiles after continuum subtraction. The latter blend probes the 
irradiating continuum and demonstrates that the X-ray ionization flux virtually vanishes after 
MJD 57200 \citep{munoz16,mata18}.  

 \begin{table}
	\centering
	\caption{Emission line fluxes during the nebular phase and decay$^{\dagger}$.}
	\label{tab:tab2}
	\begin{tabular}{lcccc}
		\hline
MJD & \multicolumn{2}{c}{Flux H$_{\alpha}$} & Flux H$_{\beta}$ & Flux Bowen \\
     & nebula & disc  &  & +HeII $\lambda$4686 \\ 
 		\hline
 57200.60   & 1140 $\pm$ 3 &  - &                     363.1  $\pm$  0.3 &  479.2  $\pm$  0.4  \\
 57200.77$^{\ast}$   &  997 $\pm$ 5 &  - &                      206 $\pm$ 4 &  -  \\
 57201.59   & 307.0  $\pm$  0.4  & - &                     50.4  $\pm$  0.1 & 3.10  $\pm$  0.12 \\
 57202.71   & 61.0 $\pm$ 0.1  & - &                       8.66   $\pm$ 0.03  & 3.32  $\pm$ 0.06   \\
 57203.69   & 44.1 $\pm$ 0.4  & - &                       7.55   $\pm$ 0.15  & 2.14  $\pm$  0.22  \\
 57205.69   & 5.3 $\pm$ 1.4  & 11.8 $\pm$ 0.9 &  4.62   $\pm$ 0.07  & 1.92   $\pm$ 0.12 \\
 57208.70   & 0.8 $\pm$ 1.4  & 6.1 $\pm$ 1.3 & - & - \\
 57211.65   & 0.3 $\pm$ 0.7  & 3.3 $\pm$ 0.8 & 0.98  $\pm$  0.06 &  $<0.30$  \\
 57216.65  & - &   2.83 $\pm$ 0.04 &                      0.82   $\pm$ 0.02 &    $<0.10$  \\
 57222.60  & - &   1.97 $\pm$ 0.07 &                      0.52   $\pm$ 0.05 &    $<0.24$  \\
 57223.64  & - &   2.75 $\pm$ 0.02 &                      0.77   $\pm$ 0.02 &    $<0.10$  \\
 57224.62  & - &   2.66 $\pm$ 0.09 &                      0.91   $\pm$ 0.10 &    $<0.54$  \\
 57225.64  & - &   2.16 $\pm$ 0.04 &  - & - \\
 57235.65  & - &   1.37 $\pm$ 0.01 &  - & - \\ 
 57267.46   & - &  1.27 $\pm$ 0.13 &                      0.23  $\pm$  0.03 &    $<0.18$  \\
\hline
	\end{tabular}
$^{\dagger}${~Fluxes are given in units of 10$^{-13}$ erg cm$^{-2}$ s$^{-1}$.}	\\
$^{\ast}${~From Rahoui et al. (2017), dereddened by $A_{V}=4.0$.}	
\end{table}

\subsection{Outflow Mass}
\label{sec:mass} 

\begin{figure}
	\includegraphics[angle=-90,width=\columnwidth]{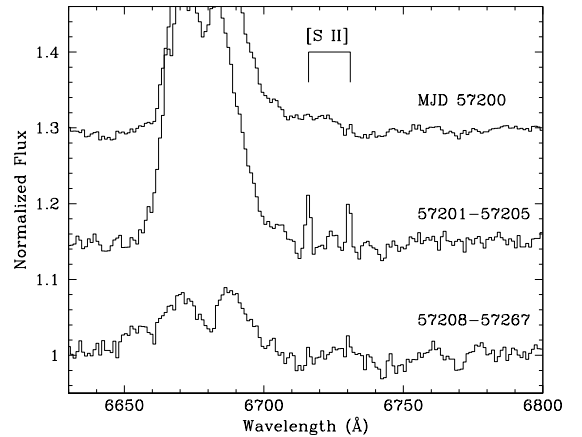}
    \caption{Example of transient forbidden lines. The $[{\rm S~II}]$ $\lambda$6717, 6731 
    doublet appears on the second day of the nebular phase and is detected throughout five 
    consecutive days (middle spectrum). } 
    \label{fig:fig4}
\end{figure}

The nebular phase in V404 Cyg has been rendered visible by a dramatic 
$\sim$3 orders of magnitude drop in X-ray luminosity in just a few hours (e.g. \citealt{maitra17}). 
The sudden {\it switch-off} of the central irradiation engine thus opens a rare opportunity to observe 
and study the glowing nebula. For example, we note the appearance of faint forbidden emission 
lines between MJD 57201 and 57205, such as $[{\rm O~III}]$ $\lambda$4959, 5007 or $[{\rm S~II}]$ 
$\lambda$6717, 6731 (see Fig~\ref{fig:fig4}). These lines are all very narrow and unresolved, with 
$FWHM$ values that are consistent with the instrumental resolution. This implies that they arise 
from a localized region, possibly an external shell, analogous to the narrow-line region in 
quasars. The intensity 
ratio of the $[{\rm S~II}]$ doublet indicates very low electron densities $N_{e} \approx10^2$ cm$^{-3}$
\citep{osterbrock89} although this diagnostic may not be reliable here because the nebula is not 
in ionization equilibrium. Furthermore, the formation site of the forbidden lines is most likely not 
representative of the bulk of the nebula. 

In order to constrain the physics of the outflow we propose instead to exploit the lifetime 
of the nebular phase as traced by the decay of the Balmer recombination flux. 
Figure~\ref{fig:fig3} shows that the \ha~nebular emission dissipates completely $\approx$10 days after 
the onset of the nebular phase. This is followed by the decay of the \ha~disc emission at a much 
slower rate. 
The evolution of the two components can be modelled by exponential decay functions with e-folding 
timescales $\tau_{\rm neb}=0.72\pm0.01$ d and  $\tau_{\rm disc}=21.4\pm0.5$ d, respectively. 
The rapid loss in \ha~luminosity is determined by the recombination timescale of the H atoms, 
$t_{\rm rec}$, as this is longer than other relevant processes such as photoionization and 
$e^{-}-e^{-}$ scattering/collisions. Therefore, we can use $t_{\rm rec} \sim (N_{e} \alpha)^{-1}$, where  
$\alpha$ is the recombination coefficient for hydrogen which, for a typical temperature $T=10^4$ K 
(as indicated by the Boltzmann plots of the nebular phase, see \citealt{mata18}) 
is $4\times10^{-13}$ cm$^3$s$^{-1}$ \citep{osterbrock89}. 
By adopting  $t_{\rm rec}=\tau_{\rm neb}$ we obtain $N_{e} \simeq5\times10^7$ cm$^{-3}$. 
 
We can now apply the relation between the intensity of the H$\beta$ emission line and the mass of hydrogen 
in the nebula \citep{osterbrock89} 

\begin{equation}
m(H)=\frac{d^2}{N_{e}} \frac{2.455\times10^{-2}}{\alpha_{H\beta}} I\left(H\beta\right)
\label{eq:mass_hb}
\end{equation}

\noindent
where $d=2.39\pm0.14$ kpc \citep{miller-jones09} and 
$\alpha_{H\beta}\approx3\times10^{-14}$ cm$^{-3}$ is 
the effective H$\beta$ recombination coefficient for $T\sim10^4$ K. From the first spectrum of the 
nebular phase on MJD 57200 we measure a de-reddened H$\beta$ flux 
$I(H\beta)=3.6\times10^{-11}$ ergs cm$^{-2}$s$^{-1}$ (see Table~\ref{tab:tab2}). Bringing this into 
equation~\ref{eq:mass_hb} yields $m(H)\simeq4\times10^{-6}$ \msun.  
Here we have taken advantage of the accurate distance  
based on the radio parallax. In the case of a 
nebula with homogeneous density this calculation is deemed to be reliable within $\sim$50\%. 
On the other hand, 
the uncertainty associated with the H$\beta$ flux is dominated by the reddening correction. For instance, 
if we adopt a lower limit $A_V=3.6$ \citep{hynes04} then $I(H\beta)$ drops by factor of 2. All in all, 
we consider that our mass calculation could be approximate to a factor $\approx$4. 

An independent estimate can be obtained if we assume that the nebula has uniform density, following

\begin{equation}
m(H)=\frac{4}{3}\pi R_{\rm neb}^{3} m_{\rm H} N_{e} \epsilon 
\label{eq:mass_radius}
\end{equation}

\noindent
where $R_{\rm neb}$ is the radius of the nebula, 
$m_{\rm H}$ the mass of a hydrogen atom 
and $\epsilon$ a term that accounts for both its filling 
factor and covering factor. 
A rough estimate of $R_{\rm neb}$ can be inferred from the mean outflow velocity,  
$v_{\rm out}\sim2000$ \kms~, as measured from the several P Cyg profiles detected during the 
outburst \citep{munoz16,mata18}. \cite{bernardini16} report a very intense 
($EW\sim300$ \AA), broad ($HWZI\sim2500$ \kms) and red-skewed \ha~profile on MJD 57188, 
indicating that significant mass outflow was already taking place $\approx$13 h before the 
X-ray trigger. Since the outburst ended abruptly on MJD 571200 \citep{ferrigno15} we then adopt  
a total outburst duration $t_{\rm out}\sim$12 d. Assuming a sustained outflow with $v_{\rm out}$ 
throughout the entire outburst we estimate the size of the nebula to be 
$R_{\rm neb}=v_{\rm out}\times t_{\rm out}\sim2\times10^{9}$ km. 
Now bringing this into equation~\ref{eq:mass_radius} and assuming $\epsilon=1$ leads to  
$m(H)\sim1\times10^{-6}$ \msun, which compares well with the previous calculation based on 
the \ha~decay. This, in turn, suggests that the nebula has large filling and covering factors. 
The latter is supported by the symmetric Gaussian profile of the nebula component, an 
indication of a spherical outflow geometry. The observation of simultaneous red-shifted emission and 
P-Cyg absorptions during the main outburst also provides independent support for a large covering 
factor \citep{munoz16}. 

It should also be mentioned that our outflow mass is consistent with that estimated from the diffusion 
timescale of the nebula, based on the transition from the optically thick P-Cyg absorption profiles into 
optically thin emission-line wings. In particular, \cite{munoz16} obtain $m(H)\sim10^{-8}-10^{-5}$ 
\msun~for diffusion timescales constrained by observations to the range 0.002 - 0.1 days. Given that the 
recombination timescale of the nebula is significantly larger than the diffusion timescale, the 
former appears as a more useful tool to estimate the mass ejected in V404 Cyg and in other systems. 

\subsection{Updated ephemeris}
\label{sec:ephemeris}

The presence of an outflow will naturally produce a change in the binary 
period that might be detected in our data (e.g. \citealt{disalvo08}). \cite{ziolkowski18} studied 
the effect of non-conservative mass-transfer in V404 Cyg, modelling the response of the 
companion's Roche lobe for the case of stripped-giant evolution and an outflow produced 
at different disc radii. They show that the 
impact on the orbital period ($P$) depends critically on where the outflow takes place. 
For an outflow located at the circularization radius $P$ is expected to increase 
at a rate $P/ \dot{P}=1.4\times10^8$ yr. If the outflow is instead 
produced at $\sim0.42 a$ (with $a$ being the binary separation i.e. close to the outer disc radius) 
it will carry away a larger fraction of specific angular momentum, resulting in 
$P/ \dot{P}=2.7\times10^7$ yr. 
 
\begin{figure}
	\includegraphics[angle=0,width=\columnwidth]{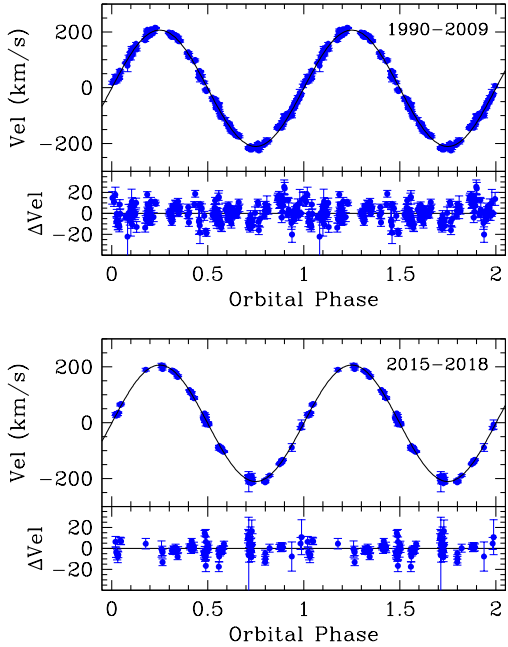}
    \caption{Heliocentric radial velocity curve of V404 Cyg on two different epochs, with the best 
    sine-wave fits overploted. Two cycles are plotted for clarity purposes. 
    The lower box on each panel shows the residuals after model subtraction.   
} 
    \label{fig:fig5}
\end{figure}

To search for possible changes in the orbital period we have revisited the orbital 
parameters of V404 Cyg. On the one hand, we have measured radial velocities from 203 quiescent 
spectra obtained between 1990 and 2009, and reported in \cite{casares94} and \cite{casares15}. 
The velocities were computed through cross-correlation with the K0 IV star template HR 8857 as in 
\cite{casares94}. 
On the other, we obtained new velocities from 80 spectra collected between 2015 and 2018 
(\citealt{mata18} and this paper). The velocities were also extracted by 
cross-correlation with the same template, broadened to $V\sin i=41$ \kms~and degraded 
to the resolution of every instrumental set-up. We have only considered the subset of high 
resolution ($<$150 \kms) spectra where the companion star is clearly detected.   
 A sine wave fit to the two groups of radial velocities results in the orbital parameters listed in the 
 first two columns of Table~\ref{tab:tab3}. 
 The systemic velocity $\gamma$ has been corrected from the radial velocity of the template, 
 that we take as -39.11$\pm$0.02 km s$^{-1}$ \citep{soubiran13}. 
 The error bars on each dataset were scaled by 
factors 2.4 and 1.8, respectively, to make the final reduced $\chi^{2}_\nu$ equal to one. All the 
quoted uncertainties in the table correspond to 1-$\sigma$. Figure~\ref{fig:fig5}  presents the 
two radial velocity curves folded on the best ephemeris. 

% \begin{landscape}
 \begin{table}
	\centering
	\caption{Orbital parameters of V404 Cyg.}
	\label{tab:tab3}
	\begin{tabular}{lccc}
		\hline
		Parameter$^{\dagger}$ & 1990-2009 & 2015-2018 & All \\
 		\hline
 P (d) &  6.471177(13) & 6.47103(16) & 6.471170(2) \\   
  T$_{0}$ (HJD-2400000) & 48089.105(5)  & 57200.518(9)  & 57200.514(2) \\
 K$_{2}$ (km s$^{-1}$)          & 208.5$\pm$0.7        &   207.8$\pm$1.5        &    208.4$\pm$0.5        \\
 $\gamma$ (km s$^{-1}$)     &  -1.0$\pm$1.0       &     -3.3$\pm$1.2        &     -2.0$\pm$0.4              \\
\hline
	\end{tabular}
$^{\dagger}${~Numbers within parenthesis indicate error in the last digits.}	
\end{table}

After comparing the two orbital solutions we conclude that it is not possible to detect a period 
variation with the present data. Despite the long baseline of the first data set, the error in $P$   
is (in the most favorable case) a factor 3 larger than the expected change. 
Therefore, even in the event of a much improved determination on the second epoch (through future 
high resolution data) it will still not be possible to measure a period variation caused by the 
2015 outflow. As pointed out by \cite{ziolkowski18}, we are limited here by the long 
orbital period of V404 Cyg. The third column in Table~\ref{tab:tab3} provides the most accurate orbital 
parameters possible, after combining all the velocities available over $\sim$30 yr.

\subsection{The Accretion Disc Evolution}
\label{sec:disc} 

In what follows, we will focus on the 
evolution of the accretion disc during the outburst decay. 
As shown by Fig.~\ref{fig:fig2}, the disappearance of the nebular emission 
gives us a glimpse at the accretion disc structure.  
In \cite{casares15} we argue that the FWHM of the \ha~line in quiescent XRTs can be taken 
as a proxy of the accretion disc radius.  In particular, inter-outburst observations of V404 Cyg obtained 
over 20 years prove the 
FWHM to be stable at 1029$\pm$94 \kms. This velocity corresponds to the Keplerian radius at 
$\sim$42\% R$_{L1}$, a value that we take as the characterisitic equilibrium radius in quiescence. 
In this context, the post-nebular evolution of the FWHM (bottom panel in Fig.~\ref{fig:fig1}) 
can be interpreted as a rapid shrinkage of the accretion disc. During this process, the disc 
is seen to overshoot the quiescent equilibrium radius before expanding back to it. 
By contrast, the disc instability theory predicts that, following the outburst expansion, the 
accretion disc should shrink slowly and steadily to the equilibrium radius because of 
the addition of low angular momentum material \citep{smak84b}. 
There is ample supporting evidence for slow disc contraction towards equilibrium in  
cataclysmic variables (e.g. \citealt{smak84a}; \citealt{odonoghue86}; \citealt{wood89};
\citealt{wolf93}; \citealt{baptista01}; \citealt{ramsay12}), and this was exactly the behaviour 
observed in the 1989 outburst of V404 Cyg as well (see Fig. 2 in \citealt{casares15}). 

\begin{figure}
	\includegraphics[angle=-90,width=\columnwidth]{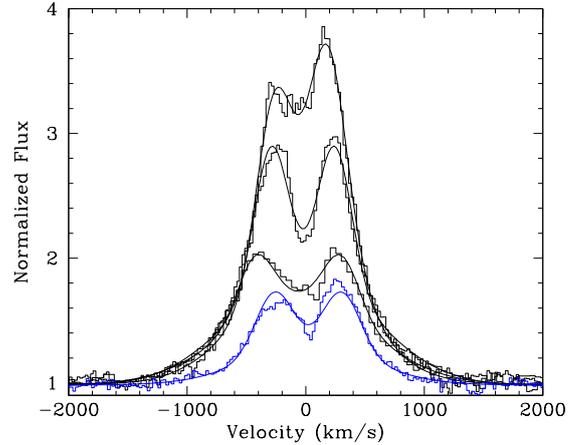}
    \caption{Widening of the \ha~double-peak separation through the outburst decay. The top three spectra 
    display disc model fits on MJD 57211, 57223 and 57235 (from top to bottom). For comparison, the 
    bottom blue spectrum depicts the quiescent average profile between MJD 57533-58430 
    (years 2016-2018).} 
    \label{fig:fig6}
\end{figure}

\begin{figure}
	\includegraphics[angle=-90,width=\columnwidth]{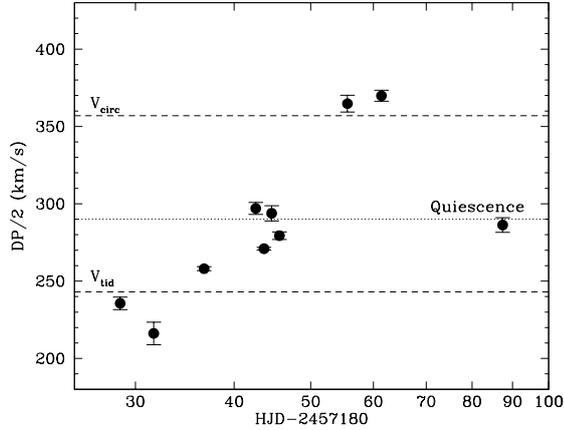}
    \caption{Time evolution of the double-peak separation of \ha~profiles. The velocities of 
    the tidal radius (${\rm V_{tid}}$), the circularization radius (${\rm V_{circ}}$) and the typical outer 
    disc radius in quiescence are indicated.}  
    \label{fig:fig7}
\end{figure}

To further investigate the puzzling evolution of the FWHM in 2015 
we have looked at the double peak separation, as delivered by our previous model fits.   
The double peak traces the outer disc velocity and, therefore, provides us with 
a direct geometrical measurement of the outer disc radius. 
Fig.~\ref{fig:fig6} presents a sample of accretion disc profiles through the outburst decay  
together with their fits,  
while Fig.~\ref{fig:fig7} plots the variation of the peak's half-separation ($DP/2$) with time.  
For reference, we have marked the disc velocities at the circularization and tidal 
truncation radii which, for the case of the V404 Cyg parameters, correspond to 
$0.37 a$ and $0.47 a$ respectively, with $a=31$ R\sun~\citep{frank02}. 
Here we have adopted $P_{\rm orb}$=6.4712 d, a binary mass ratio $q=M_{2}/M_{1}=0.067$  
\citep{casares94,casares96}, a BH mass $M_{1}$=9 \msun~and a binary inclination 
$i=67^{\circ}$  \citep{khargharia10}. Following \cite{casares16} we also assume that the 
velocity at the tidal radius is 77\% sub-Keplerian. 
In addition, we indicate the outer disc velocity in quiescence, 
as measured from the spectral average in 2016-2018 (incidentally, this is identical to the 20 year 
quiescent average obtained between the 1989 and 2015 outbursts). 

Figure~\ref{fig:fig7} confirms that the outer disc radius shrinks  
from about the tidal radius to the circularization radius in $\sim$30 days. This behaviour 
qualitatively agrees with predictions from the mass transfer instability model. For example, simulations 
by \cite{livio88} (see also \citealt{ichikawa92}) show that a burst of enhanced mass transfer triggers a 
swift disc contraction, followed by subsequent expansion  
and relaxation to the quiescent equilibrium radius. 
The contraction timescale $\Delta t$ 
reflects the inertia of the accretion disc to the addition of (low angular momentum) 
material and is largely determined by the outer disc mass $M_{\rm d}$ and the mass transfer rate 
from the donor star $\dot{M_2}$ as $\Delta t \propto M_{\rm d} / \dot{M_2}$ \citep{anderson88}. 
This property 
has been used to estimate the disc mass in some cataclysmic variables where, for typical 
values $\dot{M_2}\sim10^{-10}$ \msun~yr$^{-1}$ results in $M_{\rm d}\approx10^{-10}$ 
\msun~\citep{wood89,wolf93}. V404 Cyg has a much larger disc, with an estimated mass 
$M_{\rm d}\approx10^{-5}$ \msun~\citep{zycki99}, but the timescale of disc contraction is rather 
similar. This implies an instantaneous mass supply of $\sim10^{-5}$ 
\msun~yr$^{-1}$, that is, 4 orders of magnitude larger than the steady mass transfer rate 
expected from evolutionary considerations \citep{king93,ziolkowski18}. 

Interestingly, it has been shown that steep $\dot{M_2}$ jumps of similar amplitude 
can be produced by instabilities in X-ray illuminated atmospheres 
\citep{hameury86,hameury88,hameury90}. Hard X-rays penetrate to large Thompson optical 
depths and trigger the expansion of the sub-photospheric layers, leading to a large increase in 
mass flow through the $L_1$ point. The model presented in \cite{hameury88, hameury90} for the 
A0620-00 parameters shows an increase of $\sim$3-4 orders of magnitude from  
a quiescent value $\dot{M_{2}}\sim10^{-12}$ \msun~yr$^{-1}$.  
As a matter of fact, the amplitude of mass transfer 
enhancement is rather sensitive to the illuminating flux seen by the companion 
\citep{viallet08}. 
A detailed modelling for the specific system parameters and Eddington/super-Eddington 
hard X-ray luminosities of V404 Cyg would be required before assessing whether this is a viable 
explanation for our observations. If this were the case, we could compare our data with 
simulations performed by \cite{livio88} and \cite{ichikawa92}. These show that the disc contraction 
phase lasts until the end of the mass transfer burst. Direct scaling from our observations, 
albeit model simplifications, would imply a $\sim$30 d burst which, for the above mass transfer 
rate, yields a total of $\approx10^{-6}$ \msun~injected by the burst. 
This is comparable to the mass ejected by the outflow, although we warn that many simplifications 
are involved in this very rough estimate.  
  
\section{Discussion}
\label{sec:discussion}

V404 Cyg displays very uncommon properties among BH XRTs, with 
violent flaring activity across the entire outburst period. The outburst is further characterized by 
the presence of a sustained outflow that manifests itself through P-Cyg profiles in X-ray \citep{king15} 
and optical lines \citep{munoz16,mata18}, as well as broad emission lines (with extended wings) 
during a short-lived nebular phase
\footnote{Only V4641 Sgr has shown similar phenomenology during its brief frequent outbursts, 
with dramatic X-ray flares, optical P-Cyg absorptions and broad nebular lines (see \citealt{munoz18,chaty03}).}. 
Many papers have dealt with the timing and spectral properties of the exceptional flares 
in different energy bands 
(e.g. \citealt{rodriguez15,natalucci15,roques15,kimura15,jenke16,marti16,gandhi16,loh16,
jourdain17,rodi17,walton17,motta17a,sanchez17,maitra17,tachibana17, tetarenko17, tetarenko19}). 
Here in this paper we have focused instead on the optical study of the nebular phase and  
subsequent decay, with the aim of constraining the physics of the outflow. 

\subsection{A Massive Outflow}
\label{sec:outflow}

From the decay of the \ha~flux associated with the nebula 
we derive an outflow mass $M_{\rm wind}\simeq4\times10^{-6}$ \msun. 
This represents $\sim$10\% of the total disc mass  and a factor $\sim100$ larger 
than that accreted during the outburst, $M_{\rm acc}$  (see \citealt{zycki99}, also 
\citealt{munoz16}). The ratio 
$M_{\rm wind}/M_{\rm acc}$ implies a much larger {\it wind efficiency} than predicted by  
Compton-heated simulations \citep{done18,higginbottom19}. It also exceeds 
estimates for the soft-state wind in a sample of BH X-ray binaries, although the latter may be 
subject to large uncertainties in the ionization parameter \citep{ponti12}. 
The kinetic energy carried by the wind is 
$1/2 \dot{M}_{\rm wind} v_{\rm out}^2 \sim 4\times10^{37}$ erg s$^{-1}$, where we have adopted 
$v_{\rm out}\sim2000$ km s$^{-1}$ and assumed the outflow mass is ejected at a constant rate 
during the 12-d outburst duration. This corresponds to $\sim10^{-2} L_{\rm Edd}$, still a minor  
contribution to the energy balance but an order of magnitude larger than previous calculations   
\citep{fender16}.  

Our large wind efficiency and kinetic luminosity may be explained by the action of an extra-wind 
launching mechanism other than thermal heating. \cite{done18} and 
\cite{higginbottom19} have discussed the importance of radiation pressure at very high luminosities.  
When approaching the Eddington limit the inner disc becomes puffed up and the Compton 
radius drops substantially, allowing the wind to be launched from everywhere on the disc 
\citep{proga02}. 
This has a dramatic impact on the wind mass-loss rate and kinetic power  
and may actually represent a major contribution to the outflow during the frequent (super-)Eddington 
flares observed in V404 Cyg \citep{motta17a,motta17b}.  
Incidentally, an increase in the equivalent hydrogen absorption column $N_{\rm H}$ from 
$\sim8\times10^{21}$ cm$^{-2}$ to a few times 10$^{23}$ cm$^{-2}$ was reported during outburst 
\citep{walton15,motta17b}. This would be consistent with an outflow mass $\simeq4\times10^{-6}$ \msun, 
ejected in spherical geometry at constant $v_{\rm out}\sim2000$ \kms~over $\sim$3 days, or 
$v_{\rm out}\sim1500$ \kms~over $\sim$10 days \citep{munoz16}.

It should also be noted that our wind efficiency is beyond the critical limit 
$\dot{M}_{\rm wind}/\dot{M}_{\rm acc}\gtrsim30$ for Compton-heated winds to perturb a steady 
accretion flow and trigger $\dot{M}_{\rm acc}$ oscillations \citep{shields86}. 
In this context it is interesting to speculate whether the violent flaring activity exhibited by V404 Cyg 
is a manifestation of an accretion instability driven by the outflow. 
In any case, our large nebular mass strongly reinforces previous results on the key role played by the 
wind in the accretion/ejection balance and, even perhaps, in the occurrence of state 
transitions \citep{begelman83,neilsen09, neilsen11, ponti12}.  

\subsection{A possible mass-overflow instability}
\label{sec:mti}

Such a massive wind also poses an interesting challenge for the mass balance in the disc. 
Evolutionary considerations indicate that the donor star transfers matter at a rate $\dot{M_2}\sim10^{-9}$ 
\msun~yr$^{-1}$ \citep{king93,ziolkowski18} and, therefore, the disc is replenished with 
$\approx3\times10^{-8}$ \msun~during the 2-3 decades between outbursts. 
If $\approx10^{-6}$ \msun~are lost by the wind and a further 
$\approx10^{-8}$ \msun~through accretion \citep{zycki99,munoz16} then it seems difficult to build and 
maintain a $\approx10^{-5}$ \msun~accretion disc over $2-3$ decades unless the mass transfer 
rate from the 
companion is largely underestimated. Interestingly, the evolution of the \ha~profile shows evidence  
for rapid disc shrinkage, in support of a burst of mass transfer (Sect.~\ref{sec:disc}). The timescale of 
disc contraction suggests an extra $\approx10^{-6}$ \msun~were supplied to the disc which would be 
enough to counter wind mass losses, albeit we stress the latter figure is a crude estimation. 
In this regard, it should be noted that both the V and \ha~line 
fluxes (Figures~\ref{fig:fig1} and~\ref{fig:fig3}) decay on a $\sim$20-30 d timescale after the 
X-ray flux has faded, which is much longer than the cooling (thermal) timescale  
$t_{\rm th}\sim t_{\rm dyn}/\alpha \sim (r_{\rm circ}^{3}/G M_{1})^{0.5}/\alpha\sim$1.2 d (where 
$r_{\rm circ}$=11.4 R\sun~is the circularization radius and $\alpha \sim 0.2$ the hot-state viscosity). 
This clearly suggests that there is ongoing activity in the outer disc well after the X-ray outburst has 
ended.  Further indication is given by the fact that V stays $\sim$0.5-1 
mag above quiescence throughout the entire decay phase (Figure~\ref{fig:fig1}). 

As discussed in section~\ref{sec:disc}, a burst of enhanced mass transfer can be triggered 
by irradiation. This scenario was initially proposed by 
\cite{osaki85} to explain superoutbursts in SU UMa systems and, although it has been  
questioned for cataclysmic variables (CVs) on the basis of weak hard X-ray luminosities, it did inspire 
mass transfer instability models for BH XRTs (e.g. \citealt{chen93,augusteijn93}). 
Given the profusion of hard X-ray luminous flares detected through the 2015 outburst 
it certainly appears as a viable possibility for V404 Cyg. 

\begin{figure}
	\includegraphics[angle=-90,width=\columnwidth]{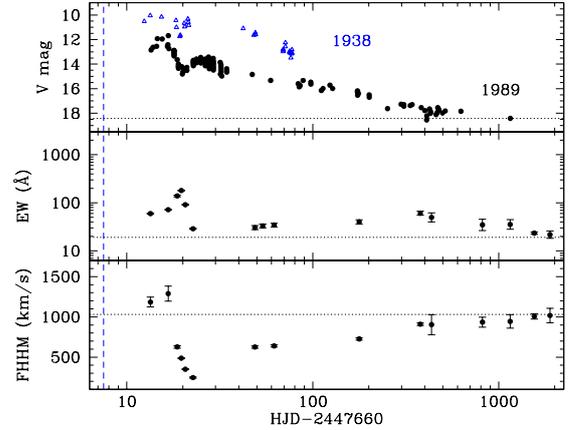}
    \caption{Same as Figure 1 but for the 1989 outburst. The blue dashed vertical line indicates the 
    {\it Ginga} trigger on 21 May 1989 (Kitamoto et al 1989). V mags are extracted from Han \& 
    Hjellming  (1992) 
    and Casares et al. (1993), while the EW and FWHM values are from Casares (2015) and Gotthelf et 
    al (1992). The first two EW/FWHM data points have been measured from digital spectra obtained 
    from Wagner et al. (1991) and Casares et al. (1991). 
    The top panel also plots in open blue triangles the photographic light curve of the historic 1938 
    outburst, from Wachmann (1948). This has been shifted vertically by -2.4 mag to prevent overlap. 
    The zero point of the 1938 light curve is taken on MJD 29174.} 
    \label{fig:fig8}
\end{figure}

\subsection{Comparison with the 1989 outburst}
\label{sec:1989}

The 2015 outburst of V404 Cyg displays striking differences with respect to previous 
episodes of activity. 
This is exemplified by Figure~\ref{fig:fig8}, where we present the long-term evolution of  
the V mag, EW and FWHM of the \ha~line for the 1989 outburst. 
The figure also plots the photographic (B-band) light curve of the 1938 outburst \citep{wachmann48}, 
when V404 Cyg was misclassified as a classical nova. Unfortunately, there is no supporting 
spectroscopic data for Nova Cygni 1938. 

A direct comparison with Figure~\ref{fig:fig1} shows that, while the 2015 outburst terminates 
$\sim$10 days after the X-ray trigger, the 1989 
(and 1938) outburst exhibit a much longer luminosity fall (also seen in radio and X-rays, e.g. 
\citealt{han92,tetarenko19}). The latter 
follows a classic exponential tail with an e-folding timescale of $\sim$110 d  that 
takes $\sim$ 2 yr to bring the system into quiescence. 
The protracted decays of the 1989 and 1938 outbursts strongly indicate that X-ray irradiation 
was key to maintain the disc in the hot state \citep{king98}. What therefore causes the abrupt 
interruption of the 2015 outburst? 

In this regard, it is interesting to note that both, the 1989 and 2015 outbursts undergo a large 
optical drop $\sim$10 days after the X-ray trigger. The downfall in 2015 is dramatic ($\sim$5 mag), 
driving the outburst to an end. Meanwhile, in 1989 the optical flux drops $\sim$2.5 mag 
and bounces back. A comparable short-lived dip and rebound is observed in the 1938 light curve 
as well. In the context of the disc instability model (DIM, see \citealt{lasota01} for a review and 
\citealt{dubus01} for detailed simulations),  a large luminosity drop 
is explained by the inward propagation of a cooling wave. 
For this to happen, the disc surface density and/or the stabilizing effect of irradiation must drop, 
allowing hydrogen to re-combine and return to the cold state. 
Alternatively, the rebound observed in 1989 requires the density behind the cooling front to be 
large enough (for a given X-ray illumination) to revive the thermal instability 
\citep{dubus99, menou00}.  

Two competing processes, external to the standard DIM although driven by consequential 
irradiation, are important actors here.  
A disc wind depletes the surface density,  
favouring the propagation of the cooling wave, while a burst of mass transfer 
temporarily enhances the outer disc mass, thus prolonging the outburst. 
The interplay between these two mechanisms, together with the irradiated flux,   
determines whether the local density lies above or below the critical limit and will 
eventually seal the fate of the outburst. 
\cite{munoz16} have proposed that the wind observed in June 2015 depletes the outer disc 
very efficiently, leading 
the outburst to a premature end. It should be noted, however, that conspicuous wind signatures in the 
form of P-Cyg absorptions were also detected in 1989 \citep{casares91}, followed by a 
clear nebular phase, with EW(\ha) peaking at 
$\sim$200 \AA~(see middle panel in Figure~\ref{fig:fig8}). On the other hand, 
\cite{mata18} noted that the P-Cyg velocities in 1989 are systematically 
lower  than in 2015, with typical values at $\sim$700 \kms. 
Lower velocities imply the wind is launched at larger disc radii, thus involving a lower 
outflow mass. In addition, the lower number of flares in 1989 (with only one briefly  
approaching Eddington on May 30, just preceding the luminosity dip; \citealt{zycki99}) 
suggest that radiation pressure was much less important. 
We here propose that the remarkable difference between the 1989 and 2015 outburst decays 
reflects different wind efficiencies and, especially, the major role played by radiation pressure in the 
latter episode. 
The abrupt end of the 2015 outburst could then be understood in terms of an effective depletion of the 
{\it inner parts} of the disc through a radiation-driven wind.  
The effect would be to quench accretion onto the BH 
and, consequently, irradiation of the outer disc, thus favouring the propagation of a cooling front. 

Another intriguing feature of the light curve is the presence of a short mini-outburst in Dec.  
2015 \citep{munoz17}.  None is seen during or immediately after the 1989 decay, although it 
should be noted that it might have been missed by sparse data coverage.   
In the aforementioned framework, it is tempting to speculate whether the Dec. 2015 mini-outburst 
could have been caused by a   
mass-transfer episode through irradiation from the final gigantic flare on MJD 57200. 
As we have discussed before, a burst of enhanced mass-loss would 
produce a disc contraction, while an accretion wave will be formed. 
After some time, the disc relaxes back to the equilibrium radius but the accretion wave 
continues travelling inwards until it is accreted by the central object. 
This scenario would be similar to the model proposed by \cite{augusteijn93}, where 
X-ray {\it glitches} are interpreted as echoes of the 
main outburst. The Dec 2015 mini-outburst occurs $\sim$150 days after the peak of the main 
outburst, a timescale that must reflect the diffusion time of the new material through the 
disc, plus the viscous time needed to replenish the truncated inner disc \citep{hameury97}. 
The propagation of the density wave should leave a signature in the form of changing \ha~line 
profiles that,  unfortunately, our poor spectroscopic coverage of the time preceding the mini-outburst  
prevent us from testing. Future intensive high-resolution spectroscopic monitoring of 
new V404 Cyg outbursts (and of other BH XRTs) can investigate this hypothesis.   

\section{Conclusions}
\label{sec:conclusions}

We have studied the decay of the nebular phase in the 2015 outburst of V404 Cyg and 
inferred an outflow mass of $\simeq4\times10^{-6}$ \msun. This is two orders of magnitude larger than 
the mass accreted during outburst, implying a much higher wind efficiency and kinetic energy than 
previous estimates for BH XRTs. Such a massive outflow might be explained by the contribution of a  
radiation-driven wind during the 
luminous flares (most importantly during the final flare on MJD 57199), in addition to other 
mechanisms such as classic thermal heating. 
This brings new support to the view that the 2015 outburst (as opposed to that in 1989)  
was prematurely ended by the outflow sweeping up inner parts of the disc, thus quenching 
accretion onto the black hole and turning off disc irradiation. 
 
The evolution of the \ha~line profile, on the other hand, shows evidence for a rapid disc 
contraction following the nebular phase. We argue that this may be caused by an episode of 
increased mass transfer, driven by the response of the secondary star to intense X-ray illumination. 
A crude estimate, based on scaling the timescale of disc contraction with CVs, 
suggest $\approx10^{-6}$ \msun~have been added to the disc at this stage. 
This, together with the evidence for a massive wind, strengthens the importance of irradiation 
effects in modelling the outburst evolution of BH XRTs.

\section*{Acknowledgements}

JC would like to acknowledge the hospitality of the Department of Physics of the 
University of Oxford, where this work started during a sabbatical visit. 
We are grateful to J.P. Lasota for interesting discussions on the DIM. 
We also thank Rosa Clavero and other members of the IAC team of support astronomers for 
undertaking the quiescent phase observations during Service time. 
JC acknowledges support by the Leverhulme Trust through the 
Visiting Professorship Grant VP2-2015-046.  Also by the Spanish Ministry of Economy, Industry and 
Competitiveness (MINECO) under grant AYA2017-83216-P. 
TMD and MPT acknowledge support via the Ram\'on y Cajal program through fellowships 
RYC-2015-18148 and RYC-2015-17854, respectively. 
JGR acknowledges support from an Advanced Fellowship from the Severo Ochoa excellence program (SEV-2015-0548) and support from the State Research Agency (AEI) of the Spanish Ministry of Science, Innovation and Universities (MCIU) and the European Regional Development Fund (FEDER) under grant AYA2017-83383-P. PAC  acknowledges  financial  support  from the  Leverhulme  Trust through an  Emeritus Fellowship. 
DMS also acknowledges support from the ERC under the European Union's Horizon 2020 research and innovation programme (grant agreement No. 715051; Spiders). 
Partly based on observations made with the GTC and WHT telescopes under Service Time 
of Spain's Instituto de Astrof\'isica de Canarias. 
We acknowledge with thanks the variable star observations from the AAVSO International Database contributed by observers worldwide and used in this research.
MOLLY software developed by T. R. Marsh is gratefully acknowledged.

%%%%%%%%%%%%%%%%%%%%%%%%%%%%%%%%%%%%%%%%%%%%%%%%%%



%\bibliographystyle{mnras}

\begin{thebibliography}{99}

\bibitem[Alfonso-Garz\'on et al.(2018)]{alfonso-garzon18}
Alfonso-Garz\'on J. et al., 2018, A\&A, 620, A110

\bibitem[Anderson(1988)]{anderson88}
Anderson N., 1988, \apj, 325, 266

\bibitem[Augusteijn et al.(1993)]{augusteijn93}
Augusteijn T., Kuulkers E., Shaham J., 1993, A\&A, 279, L13

\bibitem[Baptista \& Catal\'an(2001)]{baptista01} 
Baptista R., Catal\'an M.S., 2001, \mnras, 324, 599   

\bibitem[Barthelmy et al.(2015)]{barthelmy15} 
Barthelmy S.D., D'Ai A., D'Avanzo P., Krimm H.A., Lien A.Y.,Marshall F.E., Maselli A., Siegel M.H., 2015, \mbox{GRB Coordinates Network}, 17929

\bibitem[Begelman et al.(1983)]{begelman83} 
Begelman M.C., McKee C.F., Shields G.A., 1983, \apj, 271, 70   

\bibitem[Bernardini \& Cackett(2014)]{bernardini14}
Bernardini F., Cackett E.M., 2014, \mnras, 439, 2771

\bibitem[Bernardini et al.(2016)]{bernardini16}
Bernardini F., Russell D.M., Shaw A.W., Lewis F., Charles P.A., Koljonen K.I.I.., Lasota J.P., Casares J., 
2016, \apj, 818, L5

\bibitem[Casares et al.(1991)]{casares91}
Casares J., Charles P.A., Jones D.H.P., Rutten R.G.M., Callanan P.J., 1991, \mnras, 250, 712

\bibitem[Casares et al.(1992)]{casares92}
Casares J., Charles P.A., Naylor T., 1992, \mbox{Nature}, 355, 614 

\bibitem[Casares et al.(1993)]{casares93}
Casares J., Charles P.A., Naylor T., Pavlenko E.P., 1993, \mnras, 265, 834 

\bibitem[Casares \& Charles(1994)]{casares94}
Casares J., Charles P.A., 1994, \mnras, 271, L5 

\bibitem[Casares (1996)]{casares96}
Casares J., 1996, in Proc. 158th Coll. of IAU 208, Astrophysics and Space
Science Library, ed. A. Evans \& H. Janet Wood (Dordrecht: Kluwer), 395 

\bibitem[Casares \& Jonker(2014)]{casares-jonker14}
Casares J., Jonker P.G., 2014, \ssr, 183, 223

\bibitem[Casares (2015)]{casares15}
Casares J., 2015, \apj, 808, 80

\bibitem[Casares (2016)]{casares16}
Casares J., 2016, \apj, 822, 99

\bibitem[Charles \& Coe(2006)]{charles06}
Charles P.A., Coe M., 2006, {\it Optical, ultraviolet and infrared observations of X-ray binaries} in 
Compact stellar X-ray sources, ed. W. Lewin \& M. van der Klis, Cambridge Astrophysics Series, 
No. 39, Cambridge University Press, p.215

\bibitem[Chaty et al.(2003)]{chaty03}
Chaty S., Charles P.A., Mart\'\i{} J., Mirabel I.F., Rodr\'\i{}guez L.F., Shahbaz T., 2003, \mnras, 343, 169

\bibitem[Chen et al.(1993)]{chen93}
Chen W., Livio M., Gehrels N., 1993, \apj, 408, L5

\bibitem[Corral-Santana et al.(2016)]{corral16}
Corral-Santana J.M., Casares J., Mu\~noz-Darias T., Bauer F.E., Mart\'\i{}nez-Pais I.G., Russell D.M., 2016, A\&A, 587, A61

\bibitem[Dallilar et al.(2018)]{dallilar18}
Dallilar Y. et al., 2018, \mbox{Science}, 358, 1299 

\bibitem[Di Salvo et al.(2008)]{disalvo08}
di Salvo T., Burderi L., Riggio A., Papitto A., Menna M.T., 2008, \mnras, 389, 1851

\bibitem[Done et al.(2018)]{done18}
Done C., Tomaru R., Takahashi T., 2018, \mnras, 473, 838

\bibitem[Dubus et al.(1999)]{dubus99}	
Dubus G., Lasota J.-P., Hameury J.-M., Charles P.A., 1999, \mnras, 303, 139

\bibitem[Dubus et al.(2001)]{dubus01}	
Dubus G., Hameury J.-M., Lasota J.-P., 2001, A\&A, 373, 251

\bibitem[EHT Collaboration et al.(2019)]{eht19}	
EHT Collaboration et al., 2019, \apj, 875, L1

\bibitem[Fender (2006)]{fender06}
Fender R.P., 2006, {\it Jets in X-ray binaries} in 
Compact stellar X-ray sources, ed. W. Lewin \& M. van der Klis, Cambridge Astrophysics Series, 
No. 39, Cambridge University Press, p.381

\bibitem[Fender \& Mu\~noz-Darias(2016)]{fender16}
Fender R.P., Mu\~noz-Darias T., 2016, \mbox{LNP}, 905, 65

\bibitem[Frank et al.(2002)]{frank02}
Frank J., King A.R., Raine D.J., 2002, Accretion Power in Astrophysics
(3rd ed.; Cambridge: Cambridge Univ. Press)

\bibitem[Ferrigno et al.(2015)]{ferrigno15}
Ferrigno C., Bozzo E., Boissay R., Kuulkers E., Kretschmar P., 2015, ATel \#7731

\bibitem[Gallo et al.(2005)]{gallo05}
Gallo E., Fender R.P., Hynes R. I., 2005, \mnras, 356, 1017

\bibitem[Gandhi et al.(2016)]{gandhi16}
Gandhi P. et al., 2016, \mnras, 459, 554 

\bibitem[Gandhi et al.(2017)]{gandhi17}
Gandhi P. et al., 2017, \mbox{NatAs}, 1, 859 

\bibitem[Gotthelf et al.(1992)]{gotthelf92}
Gotthelf E., Halpern J.P., Patterson J., Rich R.M., 1992, \aj, 103 219 

\bibitem[Hameury et al.(1986)]{hameury86}
Hameury J.M., King A.R., Lasota J.P., 1986, A\&A, 162, 71

\bibitem[Hameury et al.(1988)]{hameury88}
Hameury J.M., King A.R., Lasota J.P., 1988, A\&A, 192, 187

\bibitem[Hameury et al.(1990)]{hameury90}
Hameury J.M., King A.R., Lasota J.P., 1990, \apj, 353, 585

\bibitem[Hameury et al.(1997)]{hameury97}
Hameury J.M., Lasota J.P., McClintock J.E., Narayan R., 1997, \apj, 489, 234	

\bibitem[Han \& Hjellming(1992)]{han92}
Han X., Hjellming R.M., 1992, \apj, 400, 304

\bibitem[Higginbottom et al.(2019)]{higginbottom19}
Higginbottom N., Knigge C., Long K.S., Matthews J.H., Parkinson E.J., 2019, \mnras, 484, 4635

\bibitem[Howarth (1983)]{howarth83}
Howarth I.D., 1983, \mnras, 203, 301

\bibitem[Hynes et al.(2002)]{hynes02}
Hynes R.I., Zurita C., Haswell C.A., Casares J.,. Charles P.A., Pavlenko E.P., Shugarov S.Y., Lott D.A., 
2002, \mnras, 330, 1009

\bibitem[Hynes et al.(2004)]{hynes04}
Hynes R.I. et al., 2004, \apj, 611, L125

\bibitem[Ichikawa \& Osaki(1992)]{ichikawa92}
Ichikawa S., Osaki Y., 1992, \pasj, 44, 15

\bibitem[Jourdain et al.(2017)]{jourdain17}
Jourdain E., Roques J.-P., Rodi J., 2017, \apj, 834, 130

\bibitem[Jenke et al.(2016)]{jenke16}
Jenke P.A. et al., 2016, \apj, 826, 37

\bibitem[Khargharia et al.(2010)]{khargharia10}
Khargharia J., Froning C.S., Robinson E.L., 2010, \apj, 716, 1105

\bibitem[Kimura et al.(2015)]{kimura15}
Kimura M.. et al., 2015, \mbox{Nature}, 529, 54

\bibitem[King (1993)]{king93}
King A.R.,1993, \mnras, 260, L5

\bibitem[King \& Ritter(1998)]{king98}
King A.R. Ritter H.,1998, \mnras, 293, L42

\bibitem[King et al.(2015)]{king15}
King A.L., Miller J.M., Raymond J., Reynolds M.T., Morningstar W., 2015, \apj, 813, L37

\bibitem[Kitamoto et al.(1989)]{kitamoto89}
Kitamoto S., Tsunemi H., Miyamoto S., Yamashita K., Mizobuchi S., Nakagawa M., Dotani T., 
Makino F., 1989, \mbox{Nature}, 342, 518

\bibitem[Kong et al.(2002)]{kong02}
Kong A.K.H., McClintock J.E., Garcia M.R., Murray S.S., Barret D., 2002, \apj, 570, 227

\bibitem[Lasota(2001)]{lasota01}
Lasota J.-P., 2001, \mbox{NewARev}, 45, 449

\bibitem[Livio \& Verbunt(1988)]{livio88}
Livio M., Verbunt F., 1988, \mnras, 232, 1p

\bibitem[Loh et al.(2016)]{loh16}
Loh A. et al., 2016, \mnras, 462, L111

\bibitem[Maitra et al.(2017)]{maitra17}
Maitra D., Scarpaci J.F., Grinberg V., Reynolds M.T., Markoff S., Maccarone T.J., Hynes R.I., 
2017, \apj, 851 148

\bibitem[Marsh(1989)]{marsh89}
Marsh T.R., 1989, \pasp, 101, 1032

\bibitem[Mart\'\i{} et al.(2016)]{marti16}
Mart\'\i{} J., Luque-Escamilla P.L., Garc\'\i{}a-Hern\'andez, M.T., 2016, A\&A, 586, A58

\bibitem[Mart\'\i{}n et al.(1992)]{martin92}
Mart\'\i{}n E.L., Rebolo R., Casares J., Charles P.A., 1992, \mbox{Nature} , 358, 129

\bibitem[Mart\'\i{}n et al.(1996)]{martin96}
Mart\'\i{}n E.L., Casares J., Molaro P., Rebolo R.,  Charles P.A., 1996, \mbox{NewA} , 1, 197

\bibitem[Mata S\'anchez et al.(2018)]{mata18}
Mata S\'anchez D, Mu\~noz-Darias T., Casares J., Corral-Santana J.M., Shahbaz T., 2015, \mnras, 454, 2199

\bibitem[McClintock \& Remillard(2006)]{mcclintock06}
McClintock J.E., Remillard R., 2006, {\it Black Hole Binaries} in Compact stellar X-ray sources, ed. 
W. Lewin \& M. van der Klis, Cambridge Astrophysics Series, No. 39, Cambridge University Press, p.157

\bibitem[Menou et al.(2000)]{menou00}
Menou K., Hameury J.-M., Lasota J.-P., Narayan R., 2000, \mnras, 314, 498

\bibitem[Miller-Jones et al.(2008)]{miller-jones08}
Miller-Jones J.C.A., Gallo E., Rupen M.P., Mioduszewski A.J., Brisken W., Fender R.P., Jonker P.G.,  Maccarone T.J., 2008, \mnras, 388, 1751

\bibitem[Miller-Jones et al.(2009)]{miller-jones09}
Miller-Jones J.C.A. et al., 2009, \apj, 706, L230

\bibitem[Miller-Jones et al.(2019)]{miller-jones19}
Miller-Jones J.C.A. et al., 2019, \mbox{Nature}, 569, 374

\bibitem[Motta et al.(2017a)]{motta17a} 
Motta S.E., Kajava J.J.E., S\'anchez-Fern\'andez C., Giustini M., Kuulkers E., 2017a, \mnras, 468, 891

\bibitem[Motta et al.(2017b)]{motta17b} 
Motta S.E. et al., 2017b, \mnras, 471, 1797

\bibitem[Mu\~noz-Darias et al.(2016)]{munoz16}
Mu\~noz-Darias T. et al., 2016, \mbox{Nature}, 534, 75

\bibitem[Mu\~noz-Darias et al.(2017)]{munoz17}
Mu\~noz-Darias T. et al., 2017, \mnras, 465, L124

\bibitem[Mu\~noz-Darias et al.(2018)]{munoz18}
Mu\~noz-Darias T., Torres M.A.P., Garcia M., 2018, \mnras, 479, 3987

\bibitem[Natalucci et al.(2015)]{natalucci15}
Natalucci L., Fiocchi M., Bazzano A., Ubertini P., Roques J.-P., Jourdain E., 2015, \apj, 813, L21

\bibitem[Neilsen \& Lee(2009)]{neilsen09}
Neilsen J., Lee J.C., 2009, \mbox{Nature}, 458, 481

\bibitem[Neilsen et al.(2011)]{neilsen11}
Neilsen J., Remillard R.A., Lee J.C., 2011, \apj, 737, 69

\bibitem[O'Donoghue(1986)]{odonoghue86}
O'Donoghue D., 1986, \mnras, 220, 23p

\bibitem[Oosterbroek et al.(1997)]{oosterbroek97}
Oosterbroek T. et al., 1997, A\&A, 321, 776

\bibitem[Osterbrock(1989)]{osterbrock89}
Osterbrock, D.E., 1989, Astrophysics of Gaseous Nebulae and Active
Galactic Nuclei (Mill Valley: University Science Books)

\bibitem[Osaki(1985)]{osaki85}
Osaki Y., 1985, A\&A, 144, 369

\bibitem[Piano et al.(2017)]{piano17}
Piano G., Munar-Adrover P., Verrecchia F., Tavani M.,Trushkin S.A, 2017, \apj, 839, 84

\bibitem[Plotkin et al.(2017)]{plotkin17}
Plotkin R.M. et al., 2017, \apj, 834, 104

\bibitem[Plotkin et al.(2019)]{plotkin19}
Plotkin R.M., Miller-Jones J.C.A., Chomiuk L., Strader J., Bruzewski S., Bundas A., Smith K.R., 
Ruan J.J., 2019, \apj, 874, 13

\bibitem[Ponti et al.(2012)]{ponti12}
Ponti G., Fender R.P., Begelman M.C., Dunn R.J.H., Neilsen J., Coriat M., 2012, \mnras, 422, L11 

\bibitem[Proga \& Kallman(2002)]{proga02}
Proga D., Kallman T.R., 2002, \apj, 565, 455 

\bibitem[Rahoui et al.(2017)]{rahoui17}
Rahoui F. et al., 2017, \mnras, 465, 4468

\bibitem[Ramsay et al.(2012)]{ramsay12}
Ramsay G., Cannizzo J.K., Howell S.B., Wood M.A., Still M., Barclay T., Smale A., 2012, \mnras, 425, 1479

\bibitem[Rana et al.(2016)]{rana16}
Rana V. et al., 2016, \apj, 821, 103

\bibitem[Rodriguez et al.(2015)]{rodriguez15}
Rodriguez J. et al., 2015, A\&A, 581, L9

\bibitem[Rodi et al.(2017)]{rodi17}
Rodi J., Jourdain E., Roques J.-P., 2017, \apj, 843, 3

\bibitem[Roques et al.(2015)]{roques15}
Roques J.-P., Jourdain E., Bazzano A., Fiocchi M., Natalucci L., Ubertini P., 2015, \apj, 813, L22

\bibitem[S\'anchez-Fern\'andez et al.(2017)]{sanchez17} 
S\'anchez-Fern\'andez C., Kajava J.J.E., Motta S.E., Kuulkers E., 2017, A\&A, 	602, 40

\bibitem[Sanwal et al.(1996)]{sanwal96}
Sanwal D., Robinson E.L., Zhang E., Colome C., Harvey P.M., Ramseyer T.F., Hellier C., Wood J.H., 
1996, \apj, 460, 437

\bibitem[Shields et al.(1986)]{shields86}
Shields G.A., McKee C., Lin D.N.C., Begelman M.C., 1986, \apj, 306, 90

\bibitem[Siegert et al.(2016)]{siegert16}
Siegert T. et al., 2016, \mbox{Nature}, 531, 341

\bibitem[Smak(1984a)]{smak84a}
Smak J., 1984, \mbox{Acta Astron.}, 34, 93

\bibitem[Smak(1984b)]{smak84b}
Smak J., 1984b, \mbox{Acta Astron.}, 34, 161

\bibitem[Soubiran et al.(2013)]{soubiran13}
Soubiran C., Jasniewicz G., Chemin L., Crifo F., Udry S., Hestroffer D., Katz D, 2013, A\&A, 552, 64

\bibitem[Tachibana et al.(2017)]{tachibana17}
Tachibana Y., Yoshii T., Hanayama H., Kawai N., 2017, \pasj, 69, 63

\bibitem[Tetarenko et al.(2017)]{tetarenko17}
Tetarenko A.J. et al., 2017, \mnras, 469, 3141

\bibitem[Tetarenko et al.(2019)]{tetarenko19}
Tetarenko A.J. et al., 2019, \mnras, 482, 3

\bibitem[Viallet \& Hameury(2008)]{viallet08}
Viallet M., Hamuery J.-M., 2008, A\&A, 489, 699

\bibitem[Wachmann(1948)]{wachmann48}
Wachmann A.A., 1948, \mbox{Erg. Astr. Nach.}, 11(5), E42

\bibitem[Wagner et al.(1991)]{wagner91}
Wagner R.M., Starrfield S.G., Howell S.B., Kreidl T.J., Bus, S. J., Cassatella A., Bertram R., Fried R., 
1991, \apj, 378, 293

\bibitem[Wagner et al.(1992)]{wagner92}
Wagner R.M., Kreidl T.J., Howell S.B., Starrfield S.G., 1992, \apj, 401, L97

\bibitem[Wagner et al.(1994)]{wagner94}
Wagner R.M., Starrfield S.G., Hjellming R.M., Howell S.B., Kreidl T.J., 1994, \apj, 429, L25

\bibitem[Walton et al.(2015)]{walton15}
Walton D.J. et al., 2015, \mbox{Astron. Telegr.}, \#7752.

\bibitem[Walton et al.(2017)]{walton17}
Walton D.J., 2017, \apj, 839, 110

\bibitem[Wolf et al.(1993)]{wolf93}
Wolf S., Mantel K.H., Horne K., Barwig H., Schoembs R., Baernbantner O., 1993, A\&A, 273, 160

\bibitem[Wood et al.(1989)]{wood89}
Wood J.H., 1989, \mnras, 239, 809

\bibitem[Zi\'olkowski \& Zdziarski(2018)]{ziolkowski18}
Zi\'olkowski J., Zdziarski A.A., 2018, \mnras, 480, 1580

\bibitem[Zurita et al.(2003)]{zurita03}
Zurita C., Casares J., Shahbaz T., 2003, \apj, 582, 369

\bibitem[Zurita et al.(2004)]{zurita04}
Zurita C., Casares J., Hynes R.I., Shahbaz T., Charles P.A., Pavlenko E.P., 2004, \mnras, 352, 877

\bibitem[Zycki et al.(1999)]{zycki99}
Zycki P.T., Done C., Smith D.A., 1999, \mnras, 	309, 561
	
\end{thebibliography}


%%%%%%%%%%%%%%%%%%%%%%%%%%%%%%%%%%%%%%%%%%%%%%%%%%

% Don't change these lines
\bsp	% typesetting comment
\label{lastpage}
\end{document}